# Key Factors of Wireless Real-Time Networks- From Dependability to Timeliness

***Authors:***

***Jeferson L. R. Souza***[△,+] and ***Frank Siqueira***[+]

[thejefecomp@neartword.com](thejefecomp@neartword.com), [{jeferson.l.r.souza,frank.siqueira}@ufsc.br]({jeferson.l.r.souza,frank.siqueira}@ufsc.br)

NeartWord[△]　　　　　　　Federal University of Santa Catarina (UFSC)[+]

White Paper　　　　　　　March 7, 2022

# Key Factors of Wireless Real-Time Networks - From Dependability to Timeliness[1]

*Jeferson L. R. Souza*[△,+] and **Frank Siqueira**[+]

[thejefecomp@neartword.com](thejefecomp@neartword.com), [{jeferson.l.r.souza,frank.siqueira}@ufsc.br]({jeferson.l.r.souza,frank.siqueira}@ufsc.br)

NeartWord[△]    Federal University of Santa Catarina (UFSC)[+]

### Abstract

*Offering support for real-time communications on top of a wireless network infrastructure is both a hot topic and still an open challenge. Wireless networks are not on the same level of safety, dependability, and timeliness observed in the wired realm, but they are evolving towards it. Instead of focusing on the results that need to be delivered, the key factors of wireless real-time networks are on the foundation of the network operation, defining their capability of being dependable, safe, and timely on their roots. IEEE 802.15.4 and ISA100.11a are part of this context, which we show how to be strengthened. From dealing with network inaccessibility to touching the needs of reliable communication protocols to ensure the safe and sound exchange of information, this white paper describes how we can go from dependability to timeliness. This is achieved by visiting the roots of the network operation for securing the provided communication service as a dependable, safe, and timely asset for industrial automation.*

---

[1] Available through a collaboration with International Society of Automation (ISA).
A special thanks to professor José Rufino from University of Lisbon (In Memoriam).
©~2022 - All rights reserved to the authors.
The content of this white paper can also be accessed on the following address: [https://blog.isa.org/key-factors-of-wireless-real-time-networks-from-dependability-to-timeliness](https://blog.isa.org/key-factors-of-wireless-real-time-networks-from-dependability-to-timeliness).

# 1 Introduction

There are moments when the occurrence of errors could not imply the observation of any kind of disturbance, which may be capable of disrupting the operation of a computer-based system. If we extrapolate such a concern to the communication network itself, we realise how important it is to deal with network errors, preventing the propagation of any type of issue to the applications and protocols executed on top of the established communication infrastructure [18, 11, 16, 17]. From the wired realm [18, 11] we have acquired fundamental lessons, emphasising the importance of securing dependability and timeliness properties during the network operation. The requirement of holding such properties in the domain of wireless communications is not different [16]. It is, indeed, essential to offer resilient real-time wireless communications in the presence of different sources of disturbances, from communication medium impairments to node mobility [13].

We all agree that dependability and timeliness are key factors for any kind of wireless real-time network, which can be established on top of any suitable standard such as the general-purpose IEEE 802.15.4 [7] or ISA100.11a [5] for industrial automation. However, the question is: how could we secure such fundamental factors? In this white paper, we visit the roots of the network operation for securing the provided network service as a dependable, safe, and timely asset for industrial automation. Instead of focusing on the results we need to deliver, we address aspects related to dependability and timeliness properties of the network operation, establishing a solid foundation to build resilient real-time wireless communications, which imposes little or no modification to the current networking standards.

The remaining content of this paper is organised as follows: Section 2 presents the system model, which is the solid foundation to resilient real-time wireless communications; Section 3 presents the dependability and timeliness concerns for designing resilient real-time communication systems, establishing the required communication properties every transmission/reception must be subjected to; Section 4 presents Wi-STARK, an state-of-the-art architecture offering a trustworthy communication service evolving the Medium Access Control (MAC) layer exposed service interface. Wi-STARK enables the design and use of different communication protocols, with distinguished set of requirements suitable to the safety-critical scenario at check. At the end, Section 5 presents the conclusions towards resilient (hard) real-time wireless communications.

# 2  System Model

The definition of a system model is a fundamental step to establish a solid ground for supporting a resilient real-time communication service over a wireless network. We have defined an innovative abstraction and communication model for resilient real-time wireless communications dubbed Wireless network Segment (WnS), which has been formally introduced in [13].

The WnS divides the network in elementary broadcast units, in which real-time communications are secured, implying the enforcement of dependability and timeliness properties of the entire communication system. All nodes communicate with each other within one-hop of distance, where the use of the WnS abstraction gets supported by the following statement:

> *If no real-time guarantees can be offered within communications at one hop of distance, no real-time guarantees can be offered within multi-hop communications at all.*

That means, any guarantee has to be secured first in the wireless space established within the broadcast domain of a WnS, prior to be extended end-to-end, across multiple WnS domains and communication hops.

The WnS formalisation is expressed by a 4-Tuple, $\text{WnS} \stackrel{def}{=} \langle X, x_m, C, W \rangle$, where $X$ is the set of wireless nodes members of the WnS; $x_m$ is the WnS coordinator, $x_m \in X$; $C$ represents a set of RF channels; and $W$ represents the set of networking access protocols utilised to perform frame transmissions. All transmissions within the WnS abstraction are subjected to network errors; the occurrence of such network errors is modelled as an omission, being an omission a network error that destroys a frame.

Figure 1 presents a conceptual graphical representation of the WnS; the ellipses provide a planar representation of the RF radiation pattern and therefore of the communication range of each node; despite the common use of circles, ellipses has been chosen to evidence irregularity of the RF radiation patterns observed in real radios, which may influence the behaviour of communication protocols [19]. The intersection between all communication ranges for a given channel $c_r \in C$ establishes the broadcast domain of the WnS (statements 4 and 5).

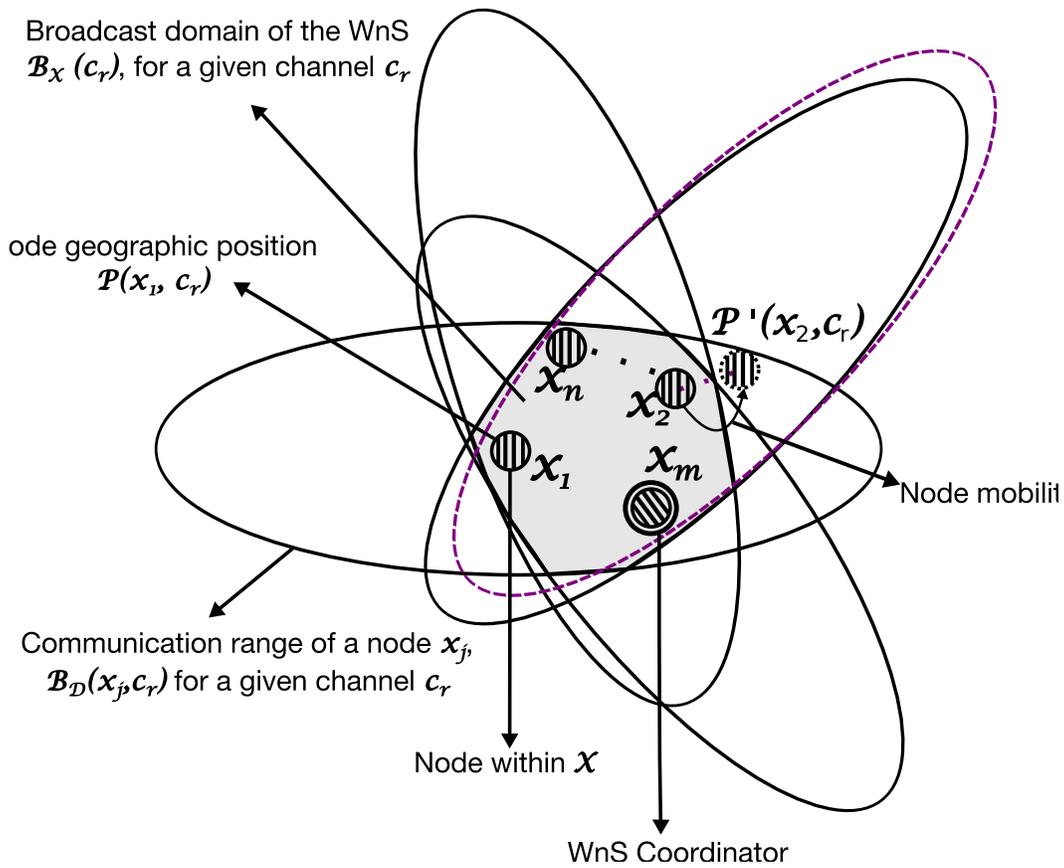

Figure 1: A conceptual graphical representation of the WnS abstraction

Each WnS is created and established by a special kind of node dubbed WnS coordinator, $x_m \in X$. The WnS coordinator is the node responsible for the management of the WnS, which includes, e.g., controlling the rules of the networking access, defining which RF channel nodes of the WnS must use, and also coordinating the entrance of new nodes in the WnS (WnS members formation).

The behaviour of the WnS abstraction is dictated by the following statements:

1. $X = \{ x_j \mid x_j$ *is a member of the WnS*$\}$ defines the set of members of the WnS (i.e., the WnS membership), where the cardinality $\#X$ represents the number of nodes that are members of the WnS, considering join and leave events that have been committed;

2. All communications are performed through a set of RF channels, $C = \{ c_r \mid c_r$ *is a RF channel used within the WnS*$\}$, where each $c_r \in C$ is a unique RF channel. The number of channels—-i.e. its cardinality—-is $\#C = f+1$, where $f$ is the number of RF communication channels that can be failed during the network operation. It is important making a distinction between the RF channel and the RF radio interface, which are related but distinguished. A RF channel represents the communication medium being in use, while the RF radio interface is the machinery enabling transmitting and/or receiving RF signals from a given RF channel;

3. All nodes $\forall x_j \in X$ use the same set of networking access protocols, which are represented by the abstract set $W = \{ w \mid w \text{ is a networking access protocol of the WnS} \}$;

4. The broadcast domain of the WnS, for a given channel $c_r \in C$, is defined by (Fig. 1):

$B_X(c_r) = \bigcap_{j=1}^{\#X} B_D(x_j, c_r)$, where $B_D(x_j, c_r)$ is a geographic region that represents the communication range of a node $x_j \in X$ for a given channel $c_r$;

5. Let $P(x_j, c_r)$ represent the geographic position of node $x_j \in X$ transmitting on channel $c_r \in C$ (Fig. 1): a node $x_j \in X$ if, and only if, $\exists c_r \in C$ where $P(x_j, c_r) \subseteq B_X(c_r)$. Otherwise, as a consequence of node's mobility, a node $x_j \notin X$ if, and only if, $\forall c_r \in C, P(x_j, c_r) \not\subseteq B_X(c_r)$;

6. Networking components (e.g., a RF channel $c_r \in C$, or a node $x_j \in X$ either behave correctly or are considered failed upon exceeding a given number of consecutive omissions (the component's *omission degree bound*), $f_o$, following a given observation criteria (e.g., the duration of a given protocol execution, $\tau_{rd}$;

7. Omissions may be inconsistent (i.e., not observed by all recipient nodes $x_j \in X$).

# 3 Dependability and Timeliness Concerns for Safety-Critical Wireless Networks

When we talk about networking communications in the domain of safety-critical environments, it is almost impossible to put aside the concerns related to the dependability and timeliness properties of communications, represented by the intrinsic requirement of a dependable and timely operation of the whole communication system. The support for such a specific and strict operation is not granted but built-in by design. In that sense, the communication characteristics of the WnS can be abstracted by a set of correctness, ordering, and timeliness properties, which are in essence independent of each particular networking technology, such as the IEEE 802.15.4 utilised by the ISA 100.11a standard. In our WnS abstraction such properties are offered through the facet of an abstract single communication channel we dubbed *WnS abstract channel* [16], as illustrated in Fig. 2. In wired communications, it has been proven that those properties are extremely useful for enforcing dependability and timeliness at higher layers protocols and applications [18, 11]. Thus, we apply a similar approach to the wireless realm as well.

Property WnS1 (*Broadcast*) formalises that it is physically impossible for a node $\texttt{x}_\texttt{j} \in X$ to send conflicting information (in the same broadcast) to different nodes, within the broadcast domain of the WnS [2], $\texttt{B}_\texttt{X}(\texttt{c}_\texttt{r})$, for a given channel $\texttt{c}_\texttt{r} \in \texttt{C}$.

Properties WnS2 (*Frame Order*) and WnS3 (*Local Full-Duplex*) are common in network technologies, wireless technologies included. Property WnS2 (*Frame Order*) is imposed by the wireless communication medium of each channel $\texttt{c}_\texttt{r} \in \texttt{C}$, and results directly from the serialisation (i.e. natural order) of frame transmissions on the shared wireless communication medium. Property WnS3 (*Local Full-Duplex*) specifies that the sender itself is also included in that ordering property, as a recipient.

Property WnS4 (*Error Detection*) has both detection and signalling facets; the detection facet, traditionally provided by the MAC layer, derives directly from frame protection through a Frame Check Sequence (FCS) mechanism, which most utilised algorithm is the cyclic redundant check (CRC) [4, 8]; the signalling facet is provided by the FCS extension introduced in [14], which is able to signal omissions detected in frames received with errors. No fundamental modifications are needed to the wireless MAC standards, such as IEEE 802.15.4 [7]. The use of such unconventional extension is enabled by emerging technologies, such as reprogrammable/reconfigurable controllers and/or open core MAC layer solutions, such as the transceivers and the MAC layers developed by ATMEL [1]. The residual probability of undetected frame errors is negligible [4, 3].

Property WnS5 (*Bounded Omission Degree*) formalises the failure semantics any entity of the WnS is subjected to, being the abstract omission degree bound, $\texttt{k} \geq \texttt{f}_\texttt{o}$. The omission degree of a WnS can be bounded, given the error characteristics of its wireless transmission medium [3, 9, 12].

The *Bounded Omission Degree* property is one of the most complex properties to secure in wireless communications. Securing this property with optimal values and with a high

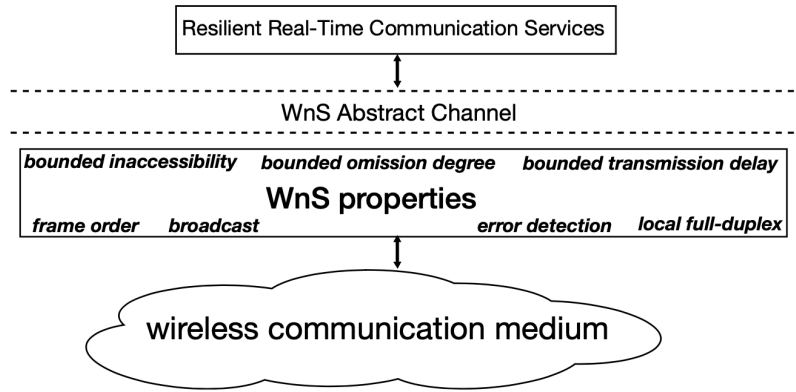

**WnS1 - *Broadcast*:** correct nodes, receiving an uncorrupted frame transmission, receive the same frame;

**WnS2 - *Frame Order*:** any two frames received at any two correct nodes are received in the same order at both nodes;

**WnS3 - *Local Full-Duplex*:** a correct node may receive, on request, local frame transmissions;

**WnS4 - *Error Detection*:** correct nodes detect and signal any corruption done during frame transmissions in a locally-received frame;

**WnS5 - *Bounded Omission Degree*:** in a known time interval $T_{rd}$, omission failures may occur in at most $k$ transmissions;

**WnS6 - *Bounded Inaccessibility*:** in a known time interval $T_{rd}$, a wireless network segment may be inaccessible at most $i$ times, with a total duration of at most $T_{ina}$;

**WnS7 - *Bounded Transmission Delay*:** any frame transmission request is transmitted on the wireless network segment, within a bounded delay $T_{td} + T_{ina}$.

Figure 2 - WnS abstract channel and its communication properties

degree of dependability coverage may require the use of multiple channels. A solution presented in [14] has shown how this can be achieved by monitoring *channel omissions*, and switch between channels upon detecting that the channel omission degree bound has been exceeded.

The behaviour of a WnS in the time domain is described by the remaining properties. Property WnS7 (*Bounded Transmission Delay*) specifies a maximum frame transmission delay, which is $T_{td}$ in the absence of faults. The value of $T_{td}$ includes the networking access and transmission delays, depending on message latency class and overall offered load bounds of real-time communication protocols, such as those specified in [10, 6]. The value of $T_{td}$ does not include the effects of omission errors. In particular, $T_{td}$ does not account for possible frame retransmissions. However, $T_{td}$ may include extra delays, e.g., resulting from longer WnS access delays derived from subtle side-effects caused by the occurrence of periods of network inaccessibility [12].

Nodes may experience a loss of connectivity within a WnS; the loss of connectivity due to transient node's mobility is also treated under the inaccessibility model. Therefore, the bounded transmission delay includes τ$_{ina}$, a corrective term that accounts for the worst-case duration of inaccessibility glitches, given the bounds specified by property WnS6 (*Bounded Inaccessibility*). The inaccessibility bounds depend on, and can be predicted by, the analysis of MAC layer characteristics [12].

# 4 The Wi-STARK Architecture for Resilient Real-Time Wireless Communications

*Wi-STARK* is a layered and component-based architecture, described in [15], which the main goal is establishing the required foundation for supporting (hard) real-time communication services on wireless realm. *Wi-STARK* has been designed in the lowest levels of the networking protocol stack, based on results obtained from a case study of the IEEE 802.15.4 standard, which can also be adapted to support ISA100.11a and other similar wireless network specifications.

The requirement of addressing the issue of network inaccessibility has been highlighted by the characterisation of network inaccessibility in IEEE 802.15.4, which has been presented in [12]. Due to the prohibitive durations obtained, which are illustrated in Fig. 3, strategies to deal with the presence of network inaccessibility in wireless realm has been designed and applied, reducing such durations significantly, as illustrated in Fig. 4.

*Wi-STARK* architecture has an intrinsic design that can be incorporated in Commercially Off-The-Shelf (COTS) components. A visual representation of the *Wi-STARK* architecture is illustrated in Fig. 5; it builds up on the RF transceiver exposed interface, which integrates \textit{basicMAC} and the Physical layer. In Fig. 5 the two distinguished but interdependent internal layers are highlighted: the **Mediator Layer** and the **Channel Layer**.

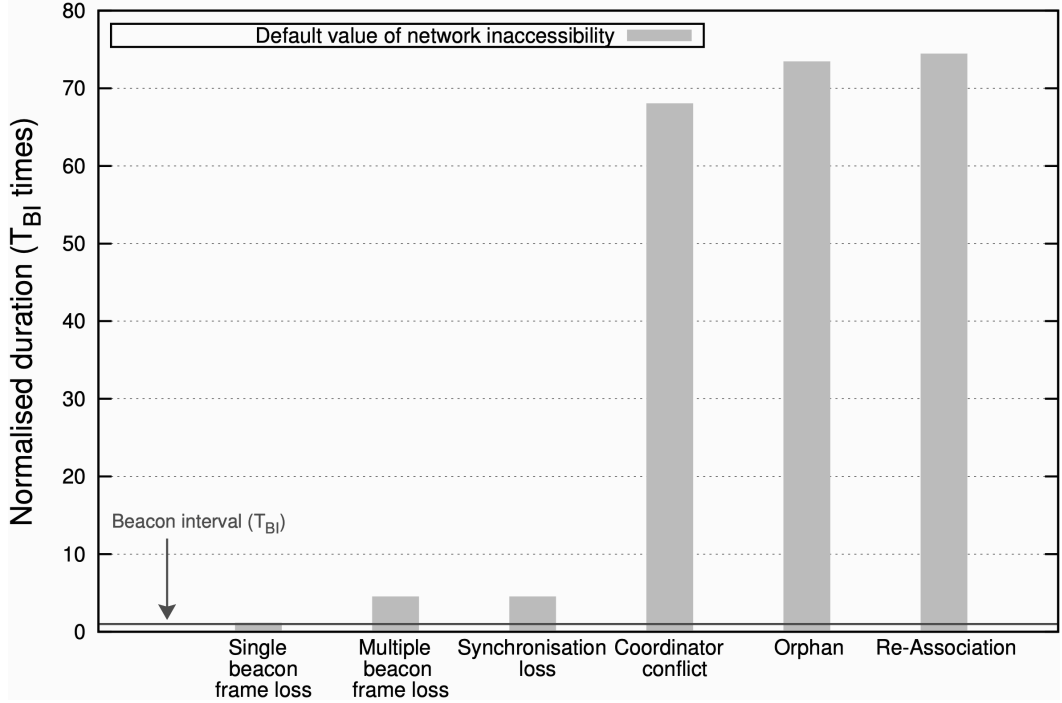

Figure 3 - Characterisation of Network Inaccessibility in IEEE 802.15.4 Wireless Networks [14]

The *Mediator Layer* is a real-time communication service layer, which enables high level entities to transmit time-sensitive data through the network. The *Channel Layer* is a control and monitoring layer, which controls and monitors the way communication channel(s) and RF transceiver(s) are utilised for real-time communication between wireless nodes. The *Wi-STARK* architecture design provides two fundamental guarantees to the high level protocol layers and applications:

**Temporal-bounded communications**: every transmitted message[2] is successfully received by all relevant correct nodes of the WnS within a known temporal bound, $\tau_{Tx\text{-}data}$.

The value of $\tau_{Tx\text{-}data}$ is directly derived from the combination of four important properties of the WnS: WnS4 (*Error Detection*), WnS5 (*Bounded Omission Degree*), WnS6 (*Bounded Inaccessibility*), and WnS7 (*Bounded Transmission Delay*). In the absence of errors, the *Wi-STARK* protocols execute in a single round, and the upper bound for all correct nodes of the WnS receiving a message successfully is: $\tau_{Tx\text{-}data}^{wc\text{-}ne} = 2.\tau_{td}$; being $\tau_{td}$ the maximum frame transmission delay in the absence of errors.

---

[2] A message is a high level protocol layer data service unit.

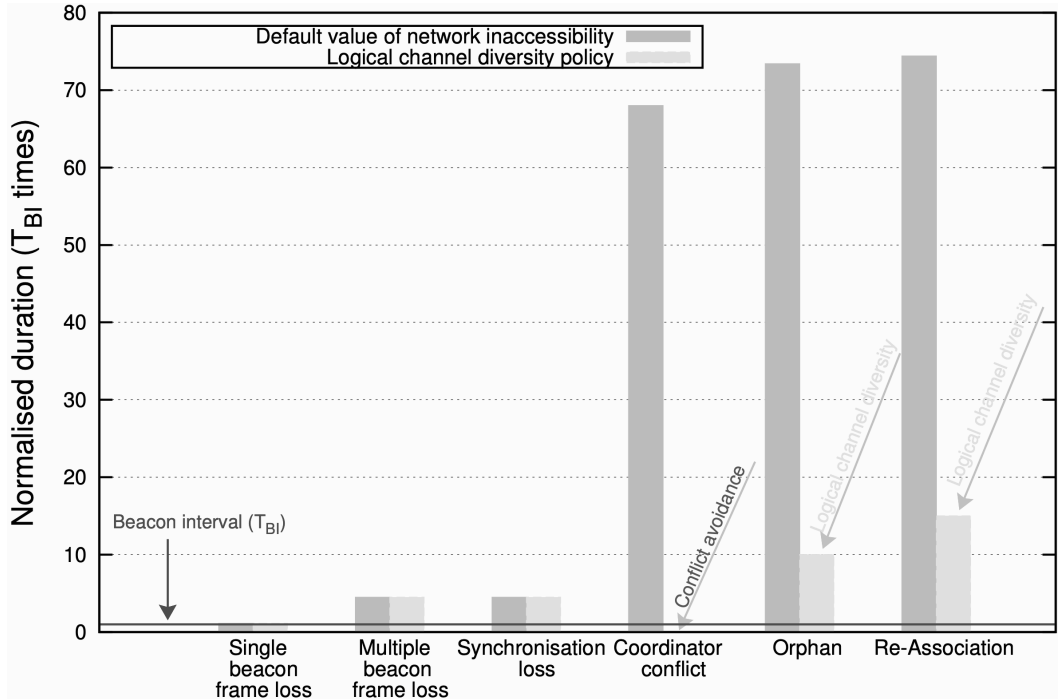

Figure 4 - Reducing Network Inaccessibility in IEEE 802.15.4 Wireless Networks [14]

In the presence of errors, frames[3] may have to be retransmitted and the protocols within the *Wi-STARK* architecture may require more than one round to be executed, up to a limit given by `k+i+1` (as specified by properties WnS5 and WnS6); all relevant correct nodes can successfully receive **any message transmitted with any reliable communication protocol}** provided by the *Wi-STARK* architecture in, at most, $\tau_{Tx\text{-}data}^{wc}$. The value of $\tau_{Tx\text{-}data}^{wc}$ depends on the protocol being executed and the behaviour of nodes, which may be either stationary, mobile, or even both. The design of such reliable communication protocol has been proposed in [16].

***Resilience against failures on the wireless communication medium***: upon the detection of a failure in the current communication channel, each correct node switches to the same communication channel within a known temporal bound of $\tau_{ina}$.

In case of communication failures, a failure of the RF communication channel in use is detected by the violation of `k`, the channel omission degree bound (WnS5), being the *Wi-*

---

[3] A frame is the MAC layer protocol data unit.

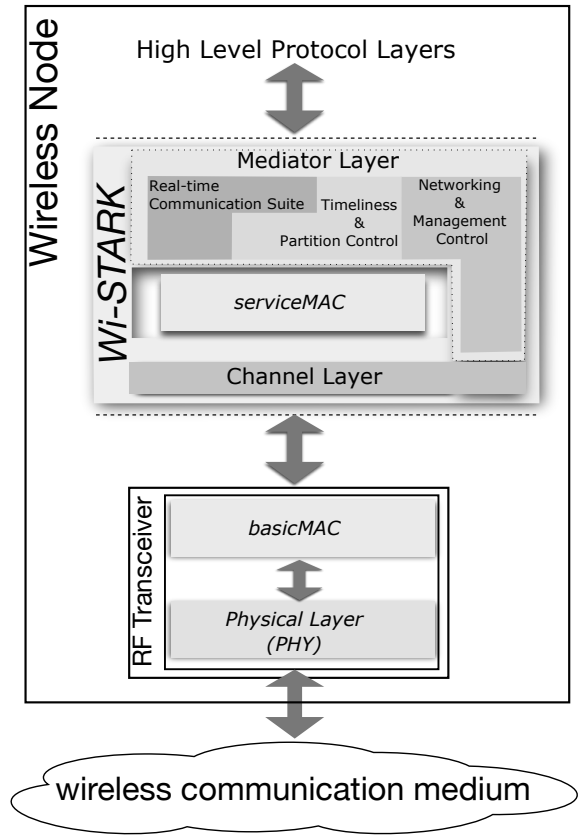

Figure 5 - The Wi-STARK Architecture

*STARK* architecture able to switch to another channel to keep the networking communications operational; the duration of the "communication blackout" resultant from that *channel failure* is then incorporated in the network inaccessibility model through $\tau_{ina}$. A more detailed description of the aforementioned channel switching strategy is described in [14].

The provision of data transmission services with temporal restrictions must take two factors into account: (a) the environment restrictions, and (b) the standard MAC and PHY layer limitations. The temporal guarantees offered by the *Real-time Communication Suite* (and the *Wi-STARK* architecture itself) fits a good balance between these two foregoing factors, e.g. utilising the maximum data rate with the minimal omission degree, $k$. It emphasises the ability to establish different levels of service, i.e. temporal guarantees, which are adaptable according to the technological limits of the MAC level standard in use.

The aforementioned levels of service are supported and tight related with the transmission protocols utilised. In essence, MAC layer provides an unreliable transmission service without delivery guarantees. Thus, the *Wi-STARK* architecture supplies a foundation to use and design protocols, which extend the aforementioned unreliable service to add delivery guarantees with an upper bounded transmission time.

In term of numbers, and evaluation of worst-case scenarios, any guarantee offered depends directly on the technology in use. As a consequence of such dependency, any guarantee offered by the *Wi-STARK* architecture is described in an abstract form, requiring a direct mapping with technology to obtain concrete numbers regarding dependability, timeliness, and safety properties of the network operation. In other words, Wi-STARK offers the capability to adapt, with little or no extension, to the characteristics of the wireless network standard in use (e.g. IEEE 802.15.4, ISA100.11a, or WirelessHART), where for each one of them we may have more strict or relaxed guarantees offered.

# 5 Conclusions

In this white paper we have addressed the key factors of wireless real-time networks, which are rested on top of a solid foundation established in the lowest levels of the wireless network protocol stack, which is able to offer dependable and timely guarantees in the network operation. We have made clear the importance of dealing with network inaccessibility through the incorporation of its side-effects into the communication properties of the WnS abstract channel, together with the pressing concern of dealing with the occurrence of network errors gracefully, and transparently, as possible to the higher layers applications and protocols executed on top the network.

By making dependability and timeliness the two fundamental pillars to achieve resilient real-time wireless communications, we step further into the realm of a more robust, trustworthy, and real-time aware wireless protocol stack, which constitutes the key for hardening the temporal guarantees offered to safety-critical environments, e.g., through a standard such as ISA100.11a.